\documentclass[aip,jcp,reprint,amsmath,amssymb,floatfix,citeautoscript]{revtex4-1}
\usepackage{cancel}
\usepackage{amsmath}
\usepackage{graphicx}
\usepackage{amssymb}
\usepackage{amsthm}
\usepackage{bm}
\usepackage{dcolumn}
\usepackage{braket}
\usepackage{longtable}
\usepackage{siunitx}

\usepackage{ragged2e}
\usepackage{txfonts}

\usepackage{color}
\usepackage[usenames,dvipsnames]{xcolor}
\definecolor{myblue}{rgb}{0,0,1}
\usepackage[breaklinks=true,colorlinks=true,linkcolor=myblue,urlcolor=myblue,citecolor=myblue]{hyperref}

\let\vr\undefined
\newcommand{\vr}{{\bm{r}}}

\newcommand{\vx}{{\bm{x}}}

\begin{document}

\title{
Quantum plasmons and intraband excitons in doped nanoparticles: Failure of the
Tamm-Dancoff approximation and importance of electron-hole attraction}

\author{Bryan T. G. Lau}
\affiliation{Department of Chemistry and James Franck Institute,
University of Chicago, Chicago, Illinois 60637, USA}
\author{Timothy C. Berkelbach}
\email{tim.berkelbach@gmail.com}
\affiliation{Department of Chemistry, Columbia University, New York, New York 10027 USA}
\affiliation{Center for Computational Quantum Physics, Flatiron Institute, New York, New York 10010 USA}

\begin{abstract}
We use excited-state quantum chemistry techniques to investigate the intraband
absorption of doped semiconductor nanoparticles as a function of doping density,
nanoparticle radius, and material properties.  The excess electrons are modeled
as interacting particles confined in a sphere.  We compare the predictions of
various single-excitation theories, including time-dependent Hartree-Fock, the
random-phase approximation, and configuration interaction with single
excitations.  We find that time-dependent Hartree-Fock most accurately describes
the character of the excitation, as compared to equation-of-motion
coupled-cluster theory with single and double excitations. The excitation
evolves from confinement-dominated, to excitonic, to plasmonic with increasing
number of electrons at fixed density, and the threshold number of electrons to
produce a plasmon increases with density due to quantum confinement.  Exchange
integrals (attractive electron-hole interactions) are essential to properly
describe excitons, and de-excitations (i.e.~avoidance of the Tamm-Dancoff
approximation) are essential to properly describe plasmons.  We propose a
schematic model whose analytic solutions closely reproduce our numerical
calculations.  Our results are in good agreement with experimental spectra of
doped ZnO nanoparticles at a doping density of $1.4\times 10^{20}$~cm$^{-3}$. 
\end{abstract}

\maketitle

\section{Introduction}

Metallic nanoparticles are an important optoelectronic platform because of their
strong plasmon resonance at visible energies, which can be tuned by size, shape,
and environment\cite{Willets2007,Odom2011}; however, the accessible carrier
densities are limited to those of the parent metals and are typically
$10^{22}$~cm$^{-3}$ or higher.  Recently, the doping of semiconductor
nanoparticles has enabled access to 
much lower electron densities~\cite{Mocatta2011,Luther2011,Scholl2012,Schimpf2014,Schimpf2015,Shen2016}. 
These doped nanoparticles, with their tunable charge carrier density, exhibit
strong intraband absorption in a wide range of the THz
regime, enabling promising new infrared and plasmonic applications~\cite{Scotognella2013,Faucheaux2014,Kriegel2017,Lhuillier2017}.

Chemical techniques have enabled semiconductor nanoparticles to be doped with
as few as 1-100 electrons. In this regime, the classical electrostatic picture of
plasmons breaks down, demanding a theory of so-called ``quantum
plasmons''~\cite{Pitarke2007,Morton2011,Bernadotte2013,Tame2013,Krauter2014,Casanova2016,Varas2016,Bursi2016,Zhang2017}.  
Furthermore, the collective plasmon picture 
becomes dubious for systems containing only 1-10 excess electrons, suggesting
a transition to ``single-particle'' excitations~\cite{Shen2016} or -- as we will argue --
excitonic transitions.
Here, we aim to present a detailed quantum mechanical understanding of the
microscopic nature of these intraband excitations over a range of
experimentally-relevant sizes and densities.

The classical Mie theory of plasmons in nanoparticles predicts a localized surface plasmon resonance when $\varepsilon_{1}(\Omega_\mathrm{LSPR})=-2\varepsilon_{\mathrm{m}}$, where
$\varepsilon_1(\omega)$ is the real part of the complex bulk dielectric function for the nanoparticle, $\varepsilon(\omega) = \varepsilon_1(\omega) + i\varepsilon_2(\omega)$
and $\varepsilon_{\mathrm{m}}$ is the dielectric constant of the medium\cite{Willets2007}.
The Drude plasmon pole approximation to the dielectric function,
$\varepsilon(\omega) 
    = \varepsilon_\infty - \Omega_\mathrm{p}^2/(\omega^2+i\gamma\omega)$,
leads to the expression for the plasmon frequency
$
\Omega_{\mathrm{LSPR}} = \sqrt{
    \Omega_\mathrm{p}^2/(\varepsilon_{\infty}+2\varepsilon_{\mathrm{m}})-\gamma^2}$,
where $\Omega_\mathrm{p}=\sqrt{4\pi \rho}$ is the bulk plasma frequency, $\rho$ is the
free charge carrier density, $\varepsilon_\infty$ is the high-frequency
dielectric constant, and $\gamma$ is the scattering rate of the electrons. 
For a given material and medium, this Mie+Drude plasmon frequency only
depends on the density and not on the size, therefore failing to
account for quantum confinement effects in small nanoparticles. The Drude dielectric function can be replaced with 
a microscopic dielectric function that accounts for quantum confinement effects, but this approach typically neglects 
interparticle interactions\cite{Kraus2017,Genzel1975,Scholl2012,Jain2014,Schimpf2014}. 

The prevailing quantum mechanical theory of plasmons in metals is the
random-phase approximation (RPA)~\cite{Bohm1953,GellMann1957}. As a theory of the ground-state energy density
of bulk metals, the RPA famously removes the divergences encountered in
finite-order perturbation theory. As a theory of the dynamical response, the RPA
predicts the collective plasmon excitation, including its dispersion and strong
oscillator strength~\cite{PinesNozieres,GiulianiVignale}. Despite its success for simple bulk metals, the RPA (by which
we mean the time-dependent Hartree approximation) is not
an accurate theory of excitation energies in molecules, casting doubt
on its applicability to quantum plasmonics. In particular, the RPA fails to
describe bound states such as excitons. To go beyond the RPA requires the tools
of higher-level many-body theory or quantum chemistry.

In this paper, we investigate a model of interacting electrons confined to the interior of a sphere of radius $R$.  This model generalizes the uniform electron gas
(UEG), sometimes referred to as ``jellium'', which is the canonical model of
bulk metals and their plasmonic excitations. In the $R\rightarrow \infty$ limit,
our model approaches the UEG (up to a background charge density).  At finite
$R$, quantum confinement produces a one-particle spectrum that is gapped, which
alters the nature of the dominant excitations. We develop the requisite
machinery, especially the two-electron integrals, and evaluate the performance
of the RPA, time-dependent Hartree-Fock (TDHF), and configuration interaction with
single excitations (CIS), which is the Tamm-Dancoff approximation (TDA) to TDHF.
For accessible system sizes, we compare our results to higher-level 
equation-of-motion coupled-cluster theory with single and double excitations
(EOM-CCSD)~\cite{Emrich1981,Koch1990,Bishop1991,Stanton1993,Bartlett2007,Krylov2008},
which has recently been used to characterize plasmons in the bulk
UEG~\cite{Lewis2019}.

The intraband excitation in our model evolves from confinement-dominated
and single-particle in character, to excitonic, to plasmonic, when the number of electrons is increased. 
By analyzing the underlying physics and comparing to higher-level
EOM-CCSD, we argue that the evolution of
the excitation is most accurately described by TDHF; the RPA and CIS are
distinct approximations to TDHF and are only able to correctly describe plasmons
and excitons, respectively. The transition from a confinement-dominated to
plasmon-like excitation can be driven by both the number and density of
electrons, and we find that for a fixed number of electrons, increasing the
density actually \textit{decreases} the plasmonic character of the excitation,
opposite to the prediction of noninteracting models~\cite{Jain2014}, due to
increasing quantum confinement. The combination of local density functional
theory (DFT) and the RPA, a popular technique in
literature\cite{prodan2003structural,Zhang2014,Zhang2017,Ipatov2018}, predicts results that are
similar to TDHF, but due to a cancellation of errors. 
The character
of the excitation across the entire range of number of electrons and density, in
particular the intermediate excitonic state, can only be described by properly
accounting for the attractive electron-hole (exchange) interaction, similar
to the situation in molecules or semiconductors. 

\section{Theory}
\subsection{Model}
\label{ssec:theory_model}

As a model of a doped nanoparticle, we treat the conduction band electrons as
particles in an infinitely deep spherical well, where the atomic details of the
nanoparticle are represented by the effective mass, radius, and
dielectric constant. When used, the dielectric constant approximates the 
effect of the ignored valence electrons and higher excitations, which screen 
the Coulomb interaction.  A more sophisticated model of the surface would
include a finite or stepped barrier, however we do not
expect our qualitative conclusions to be sensitive to the details of the
surface.  Furthermore, we neglect the dielectric
contrast with the environment, which alters both the single particle band gap
and the optical gap, although these two effects partially cancel in low-dimensional
semiconductors~\cite{Cho2018}. We will treat the two-body 
Coulomb interactions between the conduction
band electrons, which represents the focus of this work and goes beyond simple
models of noninteracting electrons under confinement.
We note that this ``jellium sphere'' model and various levels of theory have also been used
to describe the structure and excitations of nuclei (including the giant dipole resonance~\cite{ring2004nuclear}) 
and especially the optical properties of alkali clusters~\cite{Puska1985,Guet1992,deHeer1993,Brack1993,Koskinen1994PRB,Madjet1995,Patterson2019}.

In first quantization and atomic units, the total $N$-electron Hamiltonian is
\begin{equation}
\begin{split}
H &= \sum_{n=1}^{N} \left[-\frac{1}{2m^*}\nabla_n^2 + v(\vr_n)\right]
    + \sum_{n=1}^{N}\sum_{m<n} \frac{1}{|\vr_m-\vr_n|},
\end{split}
\end{equation}
where $m^*$ is the effective mass of the conduction band electrons and 
$v(r<R) = 0$ and $v(r\ge R) = \infty$. Although charge neutrality would imply an
additional $r$-dependent harmonic potential, here we neglect this potential
because many experimental procedures for nanoparticle doping (e.g.~photodoping
with hole scavengers~\cite{Schimpf2014}) do not preserve charge neutrality.

We use an orthogonal one-particle basis of eigenfunctions of the
one-electron part of the above Hamilonian, corresponding to the well-known
particle-in-a-sphere (PIS),
\begin{equation}
    \phi_{nlm}(r,\theta,\phi) = N_{nl}^{-1} j_{l}(\alpha_{nl}r) Y_{lm}(\theta,\phi) = R_{nl}(r) Y_{lm}(\theta,\phi),
\end{equation}
where $r$ is the radial coordinate, $\alpha_{nl} = k_{nl} /R$,
$k_{nl}$ is the $n$th zero of the spherical Bessel function $j_l(r)$, and the
normalization constant is $N_{nl} = \sqrt{R^3/2} \left|j_{l+1}(k_{nl})\right|$.
Each orbital is characterized by three quantum numbers $n,l,m$, with the limits
$n\geq 1$, $l\geq 0$, and  $m=-l,...,l$. The noninteracting orbital energies are given by
\begin{equation}
    \varepsilon_{nlm} = \frac{k_{nl}^2}{2m^*R^2},
\label{eq:oneeen}
\end{equation}
and are $(2l+1)$-fold degenerate. Unlike the hydrogen atom, the orbital energies
of the PIS are not degenerate with respect to the principle
quantum number $n$, but are $m$-fold degenerate for a given $n$ and $l$. 

In this orthogonal single-particle basis, the interacting, second-quantized
Hamiltonian is
\begin{equation}
H = \sum_{p} \varepsilon_p a_p^\dagger a_p
    + \frac{1}{2}\sum_{pqrs} \langle pq|rs\rangle a_p^\dagger a_q^\dagger
        a_s a_r,
\end{equation}
where the indices $pqrs$ run over spin-orbitals, i.e.~$p=(n,l,m,\sigma)$ 
and the two-electron integrals are given by 
\begin{equation}
\langle pq|rs \rangle
    = \int d\vx_1 \int d\vx_2 \phi_p^*(\vx_1) \phi_q^*(\vx_2) r_{12}^{-1} \phi_r(\vx_1) \phi_s(\vx_2),
\end{equation}
where $\vx=(\vr,\sigma)$ is a combined space and spin variable.

We study nanoparticles containing $N=2$, 8, 18, 32, 50, 72,
and 98 electrons, which correspond to closed-shell solutions of restrictedf
Hartree-Fock (RHF). For these system sizes, we find that the RHF solution 
occupies the orbitals 1s,1p,1d,...~up to 1$l_\mathrm{max}$, and therefore
these closed-shell fillings correspond to 
$N=2\sum_{l=0}^{l_\mathrm{max}}(2l+1)=2(l_\mathrm{max}+1)^2$, where
the factor of 2 accounts for spin.

The naive way to grow the basis set is to add PIS orbitals based on increasing
energy; however, the RHF orbitals are pure eigenfunctions of $l$, so increasing
the number of basis functions $n$ for each $l=0...l_\mathrm{max}$ is sufficient
to converge the ground state calculation. In order to capture singly excited 
states we add an
additional shell $l_\mathrm{max}+1$ based on the dipole selection rule $\Delta
l=\pm 1$. The rapidly increasing degeneracy of the basis
functions limits the number of electrons to 98, which we converge with 483 basis
functions ($n_\mathrm{max}$ of $[10,9,9,8,8,7,7,7]$ for
$l=0...l_\mathrm{max}+1=0...7$) requiring about 50 GB to store 
the two-electron integrals.

Expressions for the two-electron integrals $\langle pq|rs\rangle$, which are not
analytic but can be reduced to two-dimensional quadrature along the radial axis,
are given in the Appendix. All
electronic structure calculations are performed by defining a custom Hamiltonian
for use in the PySCF software package~\cite{Sun2018}.

\subsection{Excited states}

We focus on quantum chemical single-excitation theories due to their
favorable $O(N^4)$ scaling with system size, which makes them practical for future
atomistic studies.
Specifically, we consider excited states of the form
\begin{equation}
|\Psi_n\rangle = \sum_{ai} 
    \left[X_{ai} a_a^\dagger a_i + Y_{ai} a_i^\dagger a_a \right]
    |\Psi_0\rangle
\label{eq:singleexitedstate}
\end{equation}
where here and throughout $i,j,k,l$ and $a,b,c,d$ index occupied and unoccupied
HF orbitals, and $X_{ai}$ and $Y_{ai}$ correspond to coefficients for the
excitation and deexcitation of an electron from orbital $i$ to $a$,
respectively. The deexcitation operator implies that the ground state
$|\Psi_0\rangle$ is potentially correlated, though unspecified.
The amplitudes
$X_{ai}$ and $Y_{ai}$  are obtained from the eigenvalue
problem~\cite{ring2004nuclear,Scuseria2008}
\begin{equation}
\label{eq:rpa}
\left(
\begin{array}{cc}
\mathbf{A}   & \mathbf{B} \\
-\mathbf{B}^* & -\mathbf{A}^*
\end{array}
\right)
\left(
\begin{array}{c}
\mathbf{X} \\
\mathbf{Y}
\end{array}
\right)
= 
\left(
\begin{array}{c}
\mathbf{X} \\
\mathbf{Y}
\end{array}
\right) \mathbf{\Omega},
\end{equation}
where $\mathbf{\Omega}$ is a diagonal matrix of excitation energies. 
The single excitation theories considered here correspond to specific
choices of the $\mathbf{A}$ and $\mathbf{B}$ matrices.
The most ``complete'' theory is TDHF, for which
\begin{subequations}
\begin{align}
A_{ai,bj} &= (\varepsilon_a - \varepsilon_i) \delta_{ab}\delta_{ij} + \langle ib || aj \rangle, \\ 
B_{ai,bj} &= \langle ij || ab \rangle,
\end{align}
\end{subequations}
and the antisymmetrized integrals are defined as $\langle pq||rs\rangle =
\langle pq|rs\rangle - \langle pq|sr\rangle$. The RPA is obtained by neglecting
this antisymmetrization (consistent with time-dependent Hartree theory), and retains only 
the ``direct'' Coulomb matrix elements.
The neglected ``exchange'' Coulomb matrix elements (also sometimes called ``direct electron-hole
interactions'') are responsible for the
electron-hole attraction and the formation of bound excitons.
The Tamm-Dancoff approximation (TDA) corresponds to $\mathbf{B}=0$, 
which neglects potential correlation in the ground state.
When applied to TDHF and the RPA, the TDA 
leads to theories we will refer to
as CIS and the RPA(TDA). 

The RPA is the minimal theory necessary for the description
of plasmons. For the UEG, the RPA predicts a plasmon dispersion $\Omega(q)$ that
has the correct
long-range limit, $\Omega(q\rightarrow 0) = \Omega_\mathrm{p}$, where
$\Omega_\mathrm{p} = \sqrt{4\pi \rho}$ is the classical plasma frequency~\cite{PinesNozieres,GiulianiVignale}. The TDA
(including CIS) predicts a collective excitation, but one whose energy
unphysically diverges in the long-range limit, 
$\Omega^\mathrm{TDA}(q\rightarrow 0) \rightarrow \infty$. 
Here, we will see the same behavior in the $R\rightarrow \infty$ limit,
highlighting the failure of the TDA for large plasmonic nanoparticles.

For comparison, we also present results obtained at lower and higher levels
of theory, where available. At a low level, we consider simply the orbital
energy differences from the noninteracting (PIS) and mean-field (HF) theories, 
i.e.~the difference in energies of the occupied and unoccupied orbitals with the largest
transition dipole matrix element.
For HF, these orbitals are always the highest occupied and lowest unoccupied molecular orbitals (HOMO and LUMO).
At a high level, we present results from 
equation-of-motion coupled-cluster theory with single and
double excitations (EOM-CCSD) up to 50 electrons. Importantly, we note that EOM-CCSD
includes ground-state correlation and excited-state double excitations,
which both contribute to screening in an effective single-excitation
theory like the Bethe-Salpeter equation. We use a basis set that adds PIS
orbitals based on increasing energy to converge the EOM-CCSD calculations.

Since we are interested in the absorption properties of doped nanoparticles, we calculate the 
dynamical polarizability
\begin{equation}
\alpha(\omega) = \sum_{m} |\langle \Psi_0|\hat{\mu}|\Psi_m\rangle|^2 
\delta(\omega - \Omega_m)
\label{eq:suscept}
\end{equation}
where $\hat{\mu} = \sum_{n=1}^{N} \hat{r}_n = \sum_{pq} r_{pq} a_p^\dagger a_q$
is the dipole operator.

\section{Results and Discussion}

We study a model nanoparticle with $m^* = 0.28$; which is characteristic
of the conduction band of ZnO, whose plasmonic properties under doping have
been experimentally studied~\cite{Schimpf2014}.  As discussed in
Sec.~\ref{ssec:theory_model}, we consider systems containing 2, 8, 18, 32, 50,
72, and 98 electrons and study experimentally 
relevant densities from $\rho=1.4\times 10^{20}$ to $1\times 10^{22}$~cm$^{-3}$,
which defines the radius $R$ for a given number of electrons $N$.  In this range,
the radii of the nanoparticles are on the order of 1-10~nm.

\subsection{Spectral properties and peak position}

\begin{figure}
\includegraphics[scale=1]{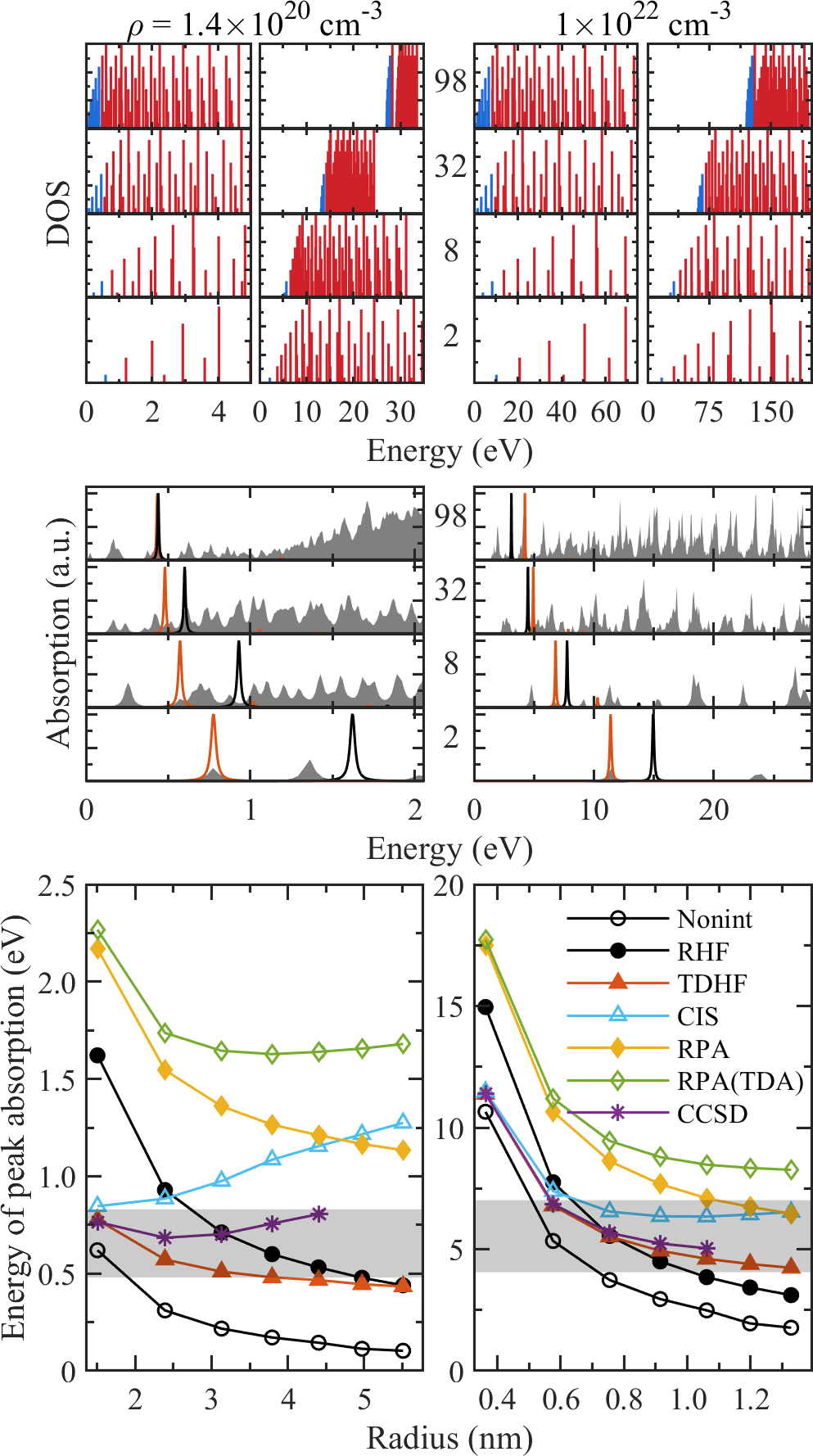}
\caption{Optical and electronic structure properties of doped nanocrystals, for
$m^* = 0.28$ at density $\rho=1.4\times 10^{20}$~cm$^{-3}$ (left) and $1\times
10^{22}$~cm$^{-3}$ (right). \textbf{Top:} The noninteracting (left column) and
mean-field (RHF, right column) density of states for 2, 8, 32, and 98 electrons.
The manifold of unoccupied states is truncated due to
the finite basis set. \textbf{Middle:} The TDHF absorption spectrum (red)
density of excited states (grey, filled); the energy of the HF gap is indicated by a black
solid peak. 
\textbf{Bottom:} Energy of
the peak absorption versus radius, for 2, 8, 18, 32, 50, 72, and 98 electrons.
The shaded region marks the bulk $\Omega_\mathrm{p}=\sqrt{4\pi \rho/m^*}$ and
Mie+Drude theory $\Omega_\mathrm{LSPR}=\Omega_\mathrm{p}/\sqrt{3}$ plasmon frequency.} 
\label{fig:1}
\end{figure}

First we analyze the single-particle orbital energies.  The top two panels of Figure~\ref{fig:1} show the the density of
states (DOS) predicted by the noninteracting (PIS, left) and mean-field (HF, right) level of theory, 
where blue lines indicate occupied orbitals and red lines indicate unoccupied orbitals.  Results are
shown for
$N=2,8,32,$ and 98 electrons (bottom to top) at the smallest and largest densities considered
here, $1.4\times 10^{20}$~cm$^{-3}$ and $1\times 10^{22}$~cm$^{-3}$ (left and right), the former
of which corresponds to the ZnO system. 
In both cases, the energy spacings are reduced with increasing $R$ due to the
scaling of the kinetic energy term, $1/(2m^*R^2)$.  In the nonineracting
results, the energy of the HOMO (roughly the Fermi energy) decreases with
increasing $R$, which is typical particle-in-a-box type physics.  However in the
HF results, the spectrum is further compressed and the energy of the HOMO is
shifted to \textit{higher} energies due to the mean-field effects of the Coulomb
interaction. 

In the middle two panels of Figure~\ref{fig:1}, we show the RHF (black) and TDHF absorption
spectrum (red), and the density of excited states $D(\omega) = \sum_m
\delta(\omega-\Omega_m)$  (grey, filled) for the same systems as above. From
the spectrum of excited states, which is becoming gapless in the limit of large
$N$, we observe a dominant peak in the absorption spectrum that arises from a
state that is typically \textit{not} the lowest in energy.  This redistribution
of oscillator strength from a low-energy continuum into a single high-energy
state is reminiscent of the plasmon peak in the dynamical structure factor of
the UEG~\cite{ring2004nuclear,Lewis2019}; the bright state is only the lowest in energy for $N=2$.
Henceforth, we focus on the excitation energy of the dominant bright peak, which
is the energy of the excited state with the largest transition dipole matrix
element.  The spectra in Figure~\ref{fig:1} also highlight a computational
challenge, as the strongly absorbing state of interest is buried in the interior
of the eigenvalue spectrum of the Hamiltonian.

In the bottom two panels of Figure~\ref{fig:1}, we compare the energy of this dominant absorption  peak versus radius, for these same two densities, as predicted by various theories.
We present results for TDHF, the RPA, CIS, RPA(TDA), and EOM-CCSD, along with
the noninteracting and HF transitions for comparison.
The EOM-CCSD result, the most accurate solution here, demonstrates the qualitative
evolution of the excitation, which we separate into three regions. At small $R$, the excitation energy
is dominated by the kinetic (confinement) energy, and scales with the band gap. At intermediate
$R$, the EOM-CCSD result is \textit{below} the HF gap and exhibits a minimum,
which we attribute to the formation of intraband excitons.  At large $R$, the
excitation energy increases with increasing $R$ and goes \textit{above} the HF
gap before reaching a plateau near the classical plasma energy.  This latter
behavior is consistent with the $R\rightarrow \infty$ limit of our model,
i.e.~the UEG, which has a well-known $q\rightarrow 0$ plasmon
at the classical plasma frequency.  

We observe from Figure~\ref{fig:1} that TDHF is the most accurate
single excitation theory. For a given density, as the
number of electrons (or radius) increases, the energy of the TDHF excitation
crosses from below to above the HF gap, which follows the EOM-CCSD result. This behavior is consistent with the physics
embodied in TDHF, which contains the ingredients necessary to capture 
the three regimes described above, i.e.~confinement-dominated excitations, excitons
(due to exchange integrals, i.e.~direct electron-hole interactions), and plasmons (due to the nonzero $\mathbf{B}$
matrix). These claims are validated by comparison with the ``approximations''
to TDHF, which contain only a subset of these ingredients: the RPA correctly predicts the evolution of the excitation to the
classical plasmon (always above the HF gap), but cannot lower the energy of 
the excitaton at any radius due
to the lack of electron-hole interactions, and CIS contains the
exchange Coulomb interaction but neglects the $\mathbf{B}$ matrix, 
so it is accurate in
describing bound excitons at small radii, but fails to correctly describe the
plasmonic state at large radii. In particular,
CIS predicts an excitation energy which goes above the HF gap and diverges at
large radius, which is precisely analogous to its $q\rightarrow 0$ behavior
in the UEG.
Finally, the RPA(TDA) result is not accurate at any radius. 

Despite the good
overall agreement of TDHF with EOM-CCSD at high electron densities, the TDHF solution suffers at low
density and large $R$ (always underpredicting the EOM-CCSD result), which is
precisely where electron correlation is expected to be strongest. 
In this regime, one may anticipate the onset of Wigner crystallization and multireference
character~\cite{Thompson2002,Jung2003,Thompson2004,Thompson2005}. While this physics could be approximately addressed via 
spin-symmetry breaking, we do not pursue this approach here.

\subsection{Characterization of the excited state}

\begin{figure}
\includegraphics[scale=1]{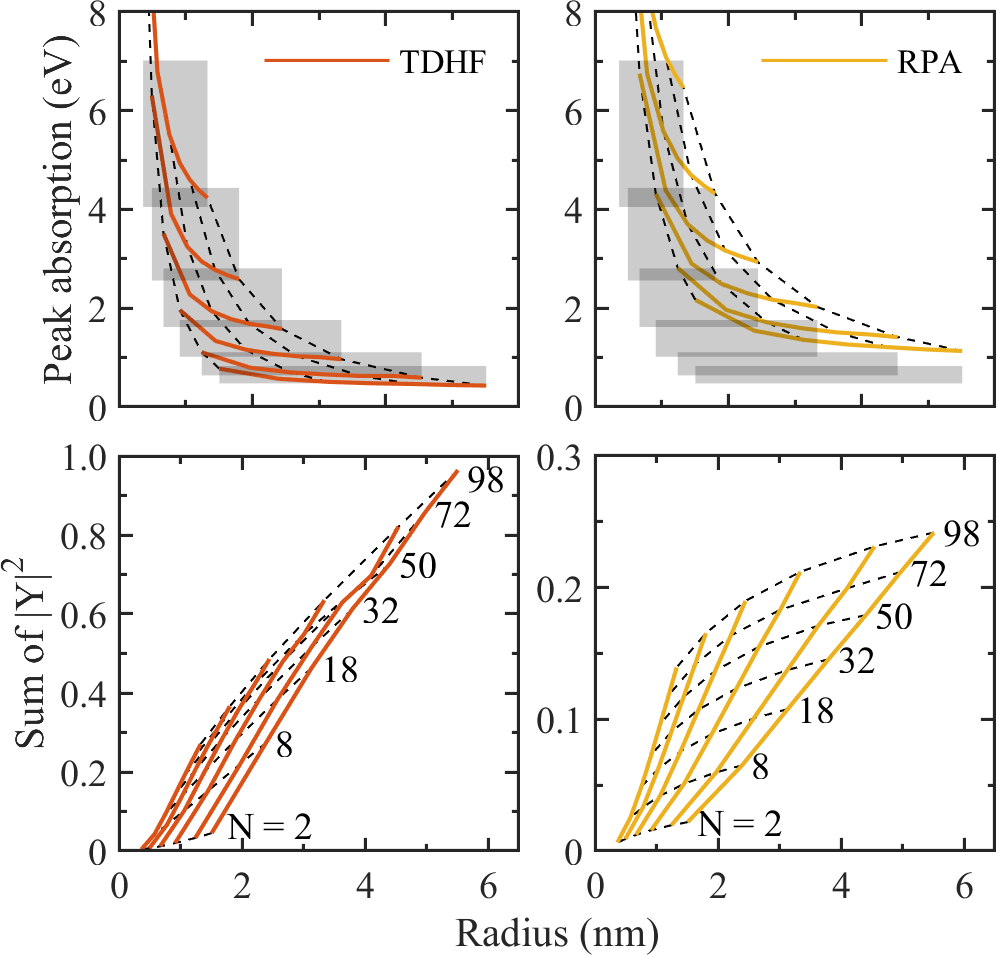}
\caption{
Summary of TDHF and RPA results for 2, 8, 18, 32, 50, 72, and 98
electrons at densities logarithmically spaced from 
\SI{1.4E20}{cm^{-3}} to \SI{1E22}{cm^{-3}}. \textbf{Top:} Energy of the peak
absorption versus radius. The shaded regions mark the bulk
$\Omega_\mathrm{p}=\sqrt{4\pi \rho/m^*}$ and Mie theory
$\Omega_\mathrm{LSPR}=\Omega_\mathrm{p}/\sqrt{3}$ plasmon frequency for each
density. Dashed lines connect results obtained with 2, 18, 50, and 98 electrons. 
\textbf{Bottom:} The sum of the deexcitation coefficients,
$\sum_{ai}|Y_{ai}|^2$, for the peak absorption state versus radius, calculated with
TDHF (left), and the RPA (right). The data are connected by lines of constant
density but changing particle number (solid) and by lines of constant particle
number but changing density (dashed).}
\label{fig:2}
\end{figure}

We now discuss a microscopic characterization of the wavefunction of the
strongly absorbing excited state. The top two plots of Figure~\ref{fig:2}
summarize the energy of the peak absorption for TDHF and the RPA, and the bottom
two plots of Figure~\ref{fig:2} plot the sum of the $Y$ coefficients for the
peak absorption, which is a measure of the plasmonic character of the
excitation. The plasmonic character is sensitive to both the absolute number of
electrons and the density, and the emergence of a plasmon is not solely a
density-driven transition. For a fixed number of electrons, the plasmonic
character decreases with increasing density, opposite to the noninteracting
result \cite{Jain2014}; at higher densities, a greater number of electrons is
needed to reach the same plasmonic character as a fewer number of electrons in a
lower density system. This decrease in plasmonic character at smaller $R$ is due
to the quantum confinement of the one-electron kinetic energy that scales as
$1/(2m^*R^2)$, and further exacerbated by the strong electron-hole interaction
(compare TDHF to the RPA). When the energy of the peak absorption is compared to
the plasmonic character, we find the surprising result that even though the
energy lies within the range of classical plasma frequencies, the excitation can
be far from plasmonic, and is instead excitonic or single-particle-like.

\begin{figure}
\includegraphics[scale=1]{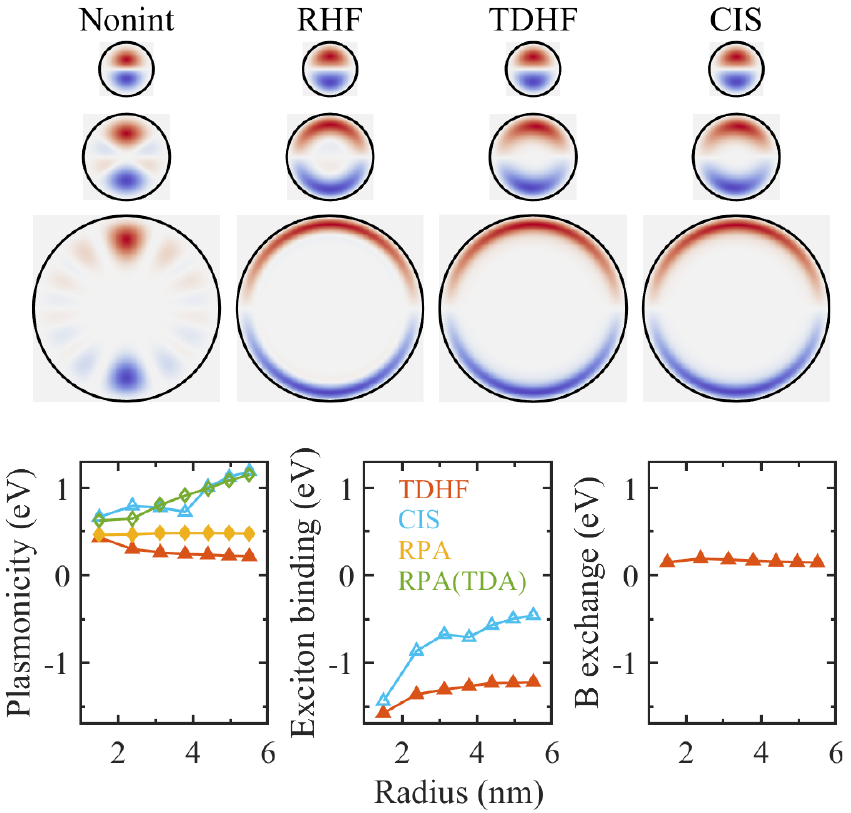}
\caption{Characterization of the strongly absorbing excited state of nanoparticles for a doping
density of density \SI{1.4e20}{cm^{-3}}. \textbf{Top:} The induced charge
density for 2 ($R=1.5$~nm), 8 ($R=2.4$~nm), and 98 electrons ($R=5.5$~nm), which
roughly correspond to a single-particle, excitonic, and plasmonic excitations.
Results are presented at four levels of theory (noninteracting, RHF, TDHF, and CIS).
\textbf{Bottom:} The many-body contributions to the energy of the peak
absorption versus radius as given in Eq.~\ref{eq:ep}.
}
\label{fig:3}
\end{figure}

Figure~\ref{fig:3} is a series of plots of
the induced charge density,
\begin{equation}
\begin{split}
\delta \rho(\vr,\omega)&=\int d\vr' \chi(\vr,\vr',\omega)v_\mathrm{ext}(\vr',\omega) \\
&= \sum_n \frac{2\Omega_n}{\omega^2-\Omega_n^2}\sum_{ai}|X_{ai}^n+Y_{ai}^n|^2\langle i|\hat{z}|a\rangle\phi_a(\vr)\phi_i(\vr),
\end{split}
\end{equation}
where $\chi$ is the density-density linear response function, for an external electric field
oriented along the $z$-axis. 
We evaluate the induced charge density at the peak absorption energy for the
noninteracting, HF, TDHF, and CIS levels of theory. The doping density of
\SI{1.4e20}{cm^{-3}} and
2 (top), 8 (middle), and 98 (bottom) electrons correspond to nanoparticle radii
of 1.5~nm, 2.4~nm, and 5.5~nm, and excitation character of confined, excitonic,
and plasmonic. As the number of electrons increases, the induced density
concentrates at the surface, in agreement with classical theory.  The addition
of mean-field interactions (HF) repels the non-interacting induced density to
the surface. The induced density at the mean-field level is not changed under configuration
mixing, because the HOMO-LUMO transition has the largest dipole matrix
element $\langle i | \hat{z} | a\rangle$ and this configuration contributes with
a large weight in the bright state.  In other words, the excited-state wavefunctions are all
qualitatively similar, despite predicting very different energies.

In light of the above observation, we next quantify the contributions of the various singles theories to
the total excitation energy,
\begin{align}
\begin{split}
\Omega_n &= 2\sum_{ai} (\varepsilon_a-\varepsilon_i) |X_{ai}+Y_{ai}|^2 \\
&\hspace{1em} +4\sum_{abij}\langle ib|aj\rangle\left(X_{bj}+Y_{bj}\right)\left(X_{ai}+Y_{ai}\right) \\
&\hspace{1em} -2\sum_{abij}\langle ib|ja\rangle\left(X_{ai}X_{bj}+Y_{ai}Y_{bj}\right) \\
&\hspace{1em} -2\sum_{abij}\langle ij|ba\rangle\left(X_{ai}Y_{bj}+Y_{ai}X_{bj}\right).
\end{split}
\label{eq:ep}
\end{align}
The second line in Eq.~(\ref{eq:ep}), containing the ``direct'' $\langle ib|aj\rangle$ integrals that describe the interaction energy between single-particle excitations, has been termed the ``plasmonicity''~\cite{Zhang2017} and is nonzero for TDHF, RPA, CIS, and RPA(TDA). The third line, containing the ``exchange'' $\langle ib|ja\rangle$ integrals from the $\mathbf{A}$ matrix, provides
a measure of the exciton binding energy and is nonzero for TDHF and CIS.  Finally, the fourth line,
containing the ``exchange'' $\langle ij|ba\rangle$ integrals from the $\mathbf{B}$ matrix has no
classical interpretation and is only nonzero for TDHF elements. 

The bottom of Figure~\ref{fig:3} plots the above contributions versus radius for a doping
density of \SI{1.4e20}{cm^{-3}}, for all four singles theories. CIS and the
RPA(TDA) do not account for ground state correlations ($\mathbf{B}=0$) and
predict a diverging plasmonicity. The RPA adds a nearly constant plasmonicity at
\textit{all} $R$, such that the excitation energy within the RPA is only modulated
by the mean-field gap, which is consistent with the prediction of a higher energy
collective excitation that is a combination of degenerate single-particle
excitations\cite{ring2004nuclear}. The RPA and RPA(TDA) overpredict the excitation
energy at small radii due to the lack of the excitonic interaction (see Figure~\ref{fig:2}, bottom). 
CIS fails at large $R$, but when quantum confinement
dominates at small $R$, both CIS and TDHF correctly describe the bound excitonic
state: notably, for the most confined case of two electrons, the exciton binding
energy lowers the excitation energy by roughly 1.5~eV. 
The exciton binding energy decreases with increasing $R$
as the electron-hole spatial overlap becomes smaller. 
The nonclassical exchange contribution from the $\mathbf{B}$ matrix makes a small
positive contribution at all values of $R$, never exceeding 0.2~eV.

TDHF mixes the correct large $R$ limit of the RPA and the small $R$ limit of
CIS. Importantly, TDHF inherits the RPA prediction of a relatively constant
plasmonicity at all radii, Figure~\ref{fig:3} bottom left. Thus, at least at this density,
there is no clearly distinguishable single-particle excited state, and the
exciton binding energy and plasmonicity make non-negligible contributions at all
values of $R$.

\subsection{Schematic model}

\begin{figure}[t]
\includegraphics[scale=1]{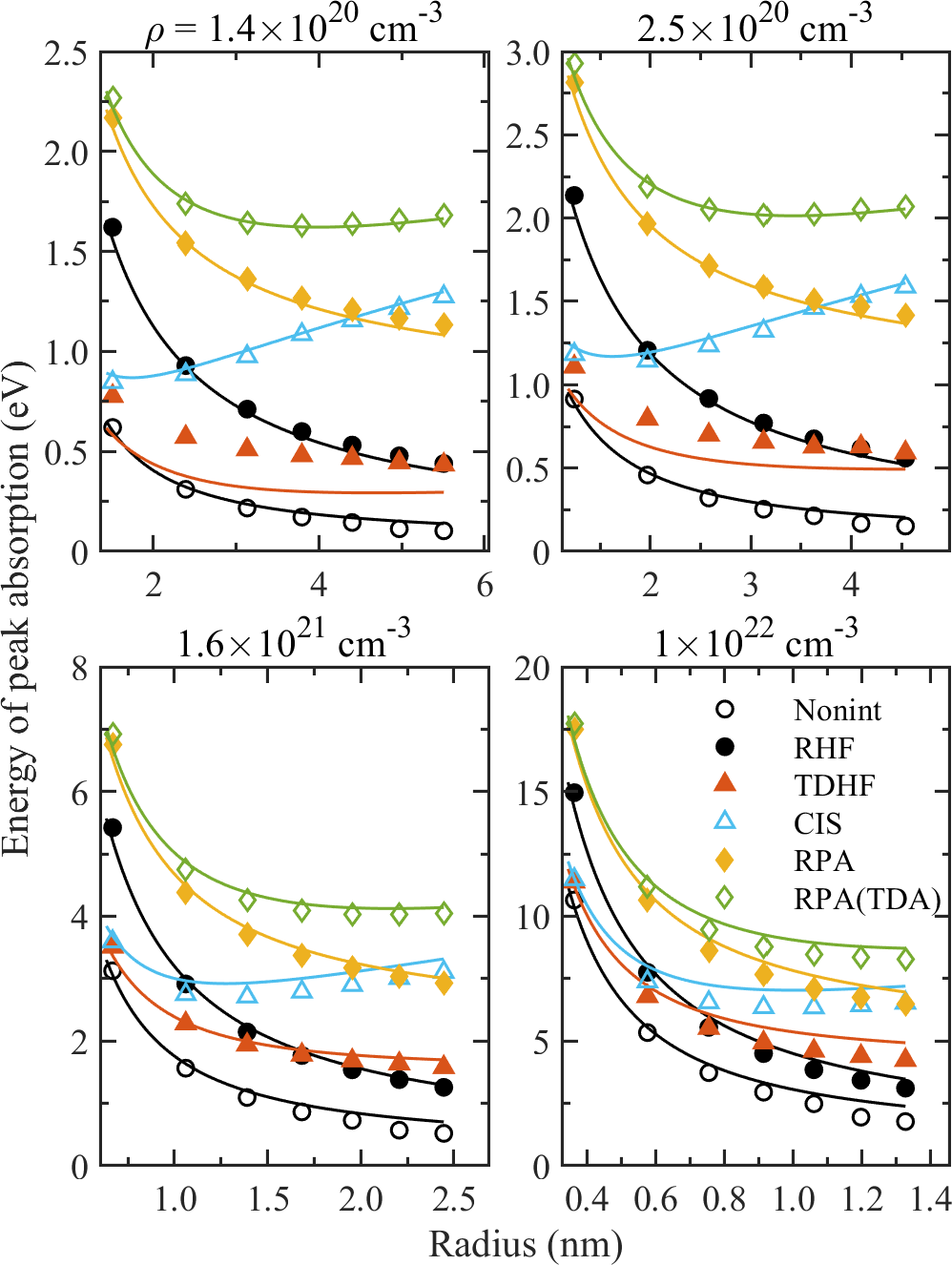}
\caption{The results of the approximate solutions of the schematic model (solid lines) compared
to the numerical results (symbols) at four different
densities.}
\label{fig:4}
\end{figure}

We can understand the dependence of the energy of the excitation on radius by
considering a schematic model of the RPA and RPA(TDA) theories, i.e.~those without
antisymmetrized integrals. This treatment is motivated by the presentation in Ref.~\onlinecite{ring2004nuclear}.
The results of the schematic model at all levels of theory will be compared to the
numerical results at four densities in Figure~\ref{fig:4}.

We consider the
factorization of the direct two-electron integrals
$\langle ib|aj\rangle \approx \lambda \rho_{ai} \rho_{bj}$, with $\lambda > 0$.
In this approximation, the RPA(TDA) equation for an excited state $n$ can be
written as
\begin{equation}
\left[ \Omega_n-(\varepsilon_a-\varepsilon_i) \right] X_{ai}^n
= \lambda \rho_{ai} \sum_{bj} \rho_{bj} X_{bj}^n,
\end{equation}
which leads to~\cite{ring2004nuclear}
\begin{align}
\frac{1}{\lambda} = \sum_{ai} 
    \frac{\rho_{ai}^2}{\Omega_n - (\varepsilon_a-\varepsilon_i)}.
\end{align}
The latter equation can be
solved graphically for $\Omega_n$, leading to a number of single-particle
excitations with energies approximately given by $\varepsilon_a - \varepsilon_i$
and one higher-energy collective excitation (the plasmon).
For illustrative purposes, we consider the subspace containing only the
HOMO and LUMO (each potentially degenerate), such that $\varepsilon_a-\varepsilon_i = \varepsilon$, which yields
for the plasmon state
\begin{align}
\Omega = \varepsilon + \lambda \sum_{ai} \rho_{ai}^2 \approx \varepsilon + \lambda N_\mathrm{trans} \bar{\rho}^2,
\end{align}
where $N_{\mathrm{trans}}$ is the number of transitions 
and $\bar{\rho}$ is an average quantity.  

We now seek to understand the $R$ dependence of this excitation energy at
\textit{fixed} density.  For all of the system sizes studied here, the
closed-shell RHF solution is only stable for those configurations for which all
occupied orbitals have $n=1$, i.e.~1s$^2$, 1p$^6$, 1d$^{10}$, 1f$^{14}$, ...,
and so on.  The angular momentum of the HOMO is thus defined by the number of
electrons, $N = 2(l_\mathrm{max}+1)^2$ or
$l_\mathrm{max} = \sqrt{N/2}-1$ at fixed density.  To a good approximation, we
find that the zeros of the spherical Bessel functions can be written as $k_{nl}
= n\pi + 1.32l$ (in particular, the value 1.32 is empirically better than the
asymptotic value $\pi/2$).  This yields a noninteracting band gap from $1l$ to
$1(l+1)$ of
\begin{equation}
\varepsilon_\mathrm{g}^{\mathrm{NI}}(R) 
    = \frac{6.55}{2m^*R^2} + \frac{5.04 \sqrt{\rho}}{2m^* \sqrt{R}}.
\end{equation} 
At the HF level, we find that the form of the one-electron
contribution to the band gap is very similar and most significantly modified by
the exchange contribution, which we model with the form $1/R$ for all densities,
$\varepsilon_\mathrm{g}^\mathrm{HF}(R) 
= \varepsilon_\mathrm{g}^{\mathrm{NI}}(R) + 1/R$.  

The HOMO and LUMO each have a degeneracy proportional
to $l(l+1)$ which, combined with the dipole selection rule
$\Delta m = 0,\pm1$, leads to a number of transitions
$N_\mathrm{trans} \propto l_\mathrm{max} \propto \sqrt{\rho R^3}$. The
Coulomb interaction has a scaling $\lambda(R) \propto R^{-1}$.
Therefore, within the RPA(TDA), the schematic model predicts an excitation
energy
\begin{equation}
\Omega^{\mathrm{RPA(TDA)}}(R)
    = \varepsilon_\mathrm{g}^{\mathrm{HF}}(R) + c \sqrt{\rho R},
\end{equation}
where we take $c=1$. At small $R$, the excitation energy of a doped
nanoparticle is given by the kinetic-energy-determined band gap; at large $R$,
the excitation energy diverges due to the Coulomb interaction. This divergence
is unphysical and analogous to the behavior of the TDA in the $q\rightarrow 0$
limit of the three-dimensional uniform electron gas.

The divergence is fixed in the full RPA, which is given in the schematic model by
\begin{align}
\frac{1}{\lambda} = \sum_{ai} \rho_{ai}^2 
    \frac{2(\varepsilon_a-\varepsilon_i)}
        {\Omega_n^2 - (\varepsilon_a-\varepsilon_i)^2}.
\end{align}
The same approximations as above for a spherical nanoparticle leads to
\begin{equation}
\Omega^{\mathrm{RPA}}(R) 
    = \sqrt{\left(\varepsilon_\mathrm{g}^\mathrm{HF}(R)\right)^2
                + 2 \sqrt{\rho R} \varepsilon_\mathrm{g}^{\mathrm{HF}}(R)}.
\end{equation}
This excitation energy has the same kinetic-energy-determined band gap at small
$R$, but now has a finite $R\rightarrow \infty$ limit, 
$\Omega^{\mathrm{RPA}}(R\rightarrow \infty)
= \sqrt{5.04\rho/m^*}$.  Importantly, the schematic model recovers the exact
limiting form of the classical plasmon energy, up to constants of order 1,
i.e.~$\sqrt{5.04} \approx 2.24$ compared to $\sqrt{4\pi}\approx 3.54$.

Before continuing on to theories with exchange (CIS and TDHF), we first 
turn to an analysis of the excitation coefficients. Again within the degenerate
schematic model, the $X$ coefficients in RPA(TDA) are
\begin{align}
    X_{ai}^{\mathrm{RPA(TDA)}} = \frac{1}{\sqrt{C}} \times \rho_{ai}
\end{align}
where $C = \sum_{ai} |\rho_{ai}|^2$ is a normalization constant,
and in the RPA are
\begin{align}
X_{ai}^{\mathrm{RPA}} &= \frac{1}{\sqrt{C}} 
    \times \frac{\rho_{ai}}{\Omega - \varepsilon_\mathrm{g}} \\
Y_{ai}^{\mathrm{RPA}} &= -\frac{1}{\sqrt{C}} 
    \times \frac{\rho_{ai}}{\Omega + \varepsilon_\mathrm{g}},
\end{align}
where $\Omega$ is the energy of the collective state and
\begin{equation}
C = 4\sum_{ai}|\rho_{ai}|^2 \varepsilon_\mathrm{g} \Omega_{\mathrm{P}}
\end{equation}
We note that as $\varepsilon_\mathrm{g}\rightarrow 0$ and the
model becomes more metallic, the excited
state is a plasmon and the $X$ and $Y$ coefficients become equal in 
magnitude.
In contrast, as the gap $\varepsilon_\mathrm{g}$ increases, the $X$
coefficients dominate and $Y_{ai}\rightarrow 0$. 
This behavior is observed numerically, as shown in Figure~\ref{fig:2}. 
The sum of deexcitation coefficients for the collective state, 
$\sum_{ai} |Y_{ai}|^2$, is therefore another measure of plasmonic character.
From Figure~\ref{fig:2}, it is clear to see that this character increases
with increasing $R$ at fixed $N$ or increasing $N$ at fixed $R$, because
both situations correspond to reducing the gap $\varepsilon_\mathrm{g}$. By this
measure, a low plasmonic character does not imply that the excitation is
single-particle-like, because the excitation can still be delocalized over the
$X_{ai}$ coefficients.

The inclusion of antisymmetrized integrals in CIS and TDHF prevents an
analytic treatment of the schematic model because the integral factorization 
$\langle ia|jb\rangle \approx \lambda \rho_{ij} \rho_{ab}$ does not facilitate
the solution of the eigenvalue problem.
However, the largest-in-magnitude element $\langle ia|ia\rangle$ can be included
exactly as it just shifts the HF band gap 
$\varepsilon_\mathrm{g}\rightarrow \varepsilon_\mathrm{g} - \langle ia|ia\rangle$.
Within the schematic model, we take all such excitonic Coulomb integrals to be equal
and obeying the scaling $\langle ia|ia\rangle = c^\prime/R$, with $c^\prime = 1.4$.
This approximation gives the CIS and TDHF excitation energies as
\begin{equation}
\Omega^{\mathrm{CIS}}(R)
    = \varepsilon_\mathrm{g}^{\mathrm{HF}}(R) + \sqrt{\rho R} - \frac{1.4}{R}
\end{equation}
\begin{equation}
\Omega^{\mathrm{TDHF}}(R) 
    = \sqrt{\left(\varepsilon_\mathrm{g}^\mathrm{HF}(R)-\frac{1.4}{R}\right)^2
                + 2 \sqrt{\rho R} \left(\varepsilon_\mathrm{g}^{\mathrm{HF}}(R)-\frac{1.4}{R}\right)}.
\end{equation}
We note that the remaining excitonic Coulomb integrals $\langle ia|jb\rangle$ could be
included via perturbation theory, though we do not pursue this here.

In Figure~\ref{fig:4}, we show the performance of the analytic solutions of this schematic model
compared to our numerical calculations. The noninteracting, HF, RPA, and
RPA(TDA) schematic models fit the data remarkably well, with the only
parameterizations being the fit of the spherical Bessel function zeros and the
proposed $1/R$ form of the HF exchange contribution to the gap. The CIS and TDHF
schematic models fit the data reasonably well, with the main difficulty being
the estimation of the contribution of the antisymmetrized two-electron
integrals. Importantly, the schematic model captures the overall behaviour of
our calculations based upon simple scaling arguments on the density and radius,
and supports the physical interpretations given to the various single-excitation theories.

\subsection{Effect of screening and comparison to DFT}

\begin{figure}
\includegraphics[scale=1]{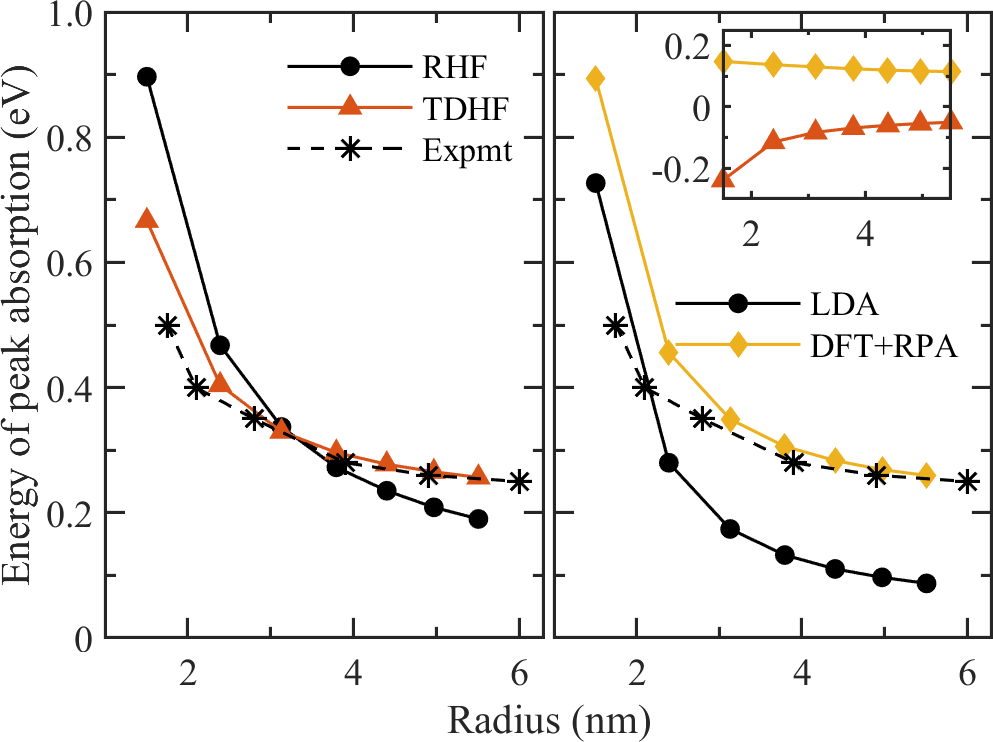}
\caption{
The energy of the peak absorption versus radius, parameterized to the
ZnO system ($\rho=1.4\times10^{20}$, $m^*=0.28$,
$\varepsilon=3.72$). \textbf{Left:} theories with
exchange: RHF gap and TDHF excitation energy
compared to the experimental results of Ref.~\onlinecite{Schimpf2014}.
\textbf{Right:} theories without exchange: LDA gap and
the DFT+RPA excitation energy, again compared to to experiment. The inset at right 
is the plasmonicity energy from TDHF (red) and DFT+RPA (yellow).}
\label{fig:5}
\end{figure}

In our model, the effect of screening from the ignored valence electrons can be approximately 
accounted for by a scaling of the Coulomb interaction,
$r_{12}^{-1}\rightarrow(\varepsilon r_{12})^{-1}$. We can approximate ZnO nanoparticles
by setting $\varepsilon=3.72$. Figure~\ref{fig:5} plots the energy of the
intraband absorption as a function of radius for this ``ZnO'' nanoparticle, at
the experimental doping density of \SI{1.4E20}{cm^{-3}}, against the
experimental results from Ref.~\onlinecite{Schimpf2014}. We compare TDHF (left) with DFT plus the RPA (right), where
the RPA excited state is calculated from the solution of the restricted
Kohn-Sham equations with the local density approximation (LDA). The LDA+RPA
approach has been applied previously to our model~\cite{Varas2016,Zhang2017} and has found success in describing the
experimental ZnO results~\cite{Goings2014,Ipatov2018,Gerchikov2018}.

From Figure~\ref{fig:5}, both TDHF and DFT+RPA compare favorably with the
experimental result, although they embody different physical effects as explored
in this paper. As usual, the LDA gap is smaller than the RHF gap, due to the
latter's treatment of exchange. The inset on the right of Figure~\ref{fig:5}
plots the sum of all contributions given in Eq.~(\ref{eq:ep}) \textit{except}
for the orbital-energy contribution, from LDA+RPA and RHF+TDHF. The RPA only acts to increase
the small LDA gap at all $R$; by contrast, the TDHF lowers the large RHF gap at
small $R$ (due to excitonic effects) and increases the RHF gap at large $R$ (due
to plasmonic effects). Therefore, ignoring exchange in the ground state and
excited state, as in the LDA+RPA approach, can produce an accurate result due to
cancellation of errors\cite{Jung2004}. 

\section{Conclusions}

In summary, we provide a fully quantum mechanical study of a confined, interacting electron gas 
as a model for doped semiconductor nanoparticles.  We observe strongly absorbing excited states
whose wavefunction character can be classified as single-particle-like (confinement dominated),
excitonic, or plasmonic.  Within the framework of the most computationally affordable single-excitation
theories, only TDHF is capable of capturing the qualitative behavior at all studied densities and particle
sizes.  We also present a schematic model of the strongly absorbing excited state that reproduces
the $R$-dependence observed in our simulations.

Our model is simple in order to focus on the essential features of electronic
interactions in the excited states of confined systems.  The model neglects atomistic details as well as surface,
ligand, or solvent effects.  The model is also ignorant of the doping mechanism and neglects the
atomic defect potential that is introduced by impurity doping (but not by electron transfer or photodoping).
Nonetheless, the results of our calculations argue strongly against the interpretation or simulation of
doped nanoparticle spectra based on single-particle transitions between orbitals, and we propose an interpretation
of intraband excitons as the primary excitations at low doping or small nanoparticles.

Looking forward towards atomistic or tight-binding~\cite{Pi2013} simulations, our work has two important ramifications. First, the
TDA fails spectacularly and should be avoided in all simulations seeking to address
the possibility of plasmonic excitations.  Second, the retention of attractive electron-hole ``exchange'' integrals
is essential for an accurate wavefunction description of excitonic states.
With these criteria in mind,
we suggest that the most promising and affordable ab initio methods are TDHF (as explored here), TDDFT with hybrid
functionals, or the $GW$+Bethe-Salpeter equation approach without the TDA~\cite{Sander2015}.

\section*{Acknowledgments}

T.C.B.~thanks Philippe Guyot-Sionnest for early conversations related to this work.
All calculations were performed with the PySCF software package~\cite{Sun2018},
using resources provided by the University of Chicago Research Computing Center.
This work was supported by the Air Force Office of Scientific Research under
AFOSR Award No.~FA9550-18-1-0058 and by the National Science Foundation CAREER
program under Award No.~CHE-1848369.  
The Flatiron Institute is a division of
the Simons Foundation.

\appendix

\section{Two-electron integrals for particle-in-a-sphere orbitals}

The spherical harmonics $Y_l^m$ are generally complex. To maximize the symmetry
of the two-electron integrals, we use the real form of the spherical harmonics,
$y_{l\mu} = \sum_{m} U_{m\mu}^{l} Y_l^m$.
With this choice, the two-electron integrals $\langle pq|rs\rangle$ are given by
\begin{equation}
\begin{split}
&\langle n_1 l_1 \mu_1; n_2 l_2 \mu_2 | n_3 l_3 \mu_3; n_4 l_4 \mu_4 \rangle \\
&= \sum_{l=0}^{\infty} R^l(n_1 l_1, n_2 l_2; n_3 l_3, n_4 l_4) 
\sum_{\mu=-l}^l \langle l\mu | l_1 \mu_1 | l_2 \mu_2\rangle \langle l\mu | l_3 \mu_3 | l_4 \mu_4\rangle,
\end{split}
\end{equation}
where the angular integrals are
\begin{equation}
\begin{split}
\langle l\mu|l_1 \mu_1|l_2 \mu_2\rangle &= \int d\Omega y_{l\mu} y_{l_1 \mu_1} y_{l_2\mu_2}\\
&=\sum_{m_1 m_2 m_3}[U_{m\mu}^{l}]^*U_{m_1 \mu_1}^{l_1}U_{m_2 \mu_2}^{l_2}\langle lm|l_1 m_1|l_2 m_2\rangle,
\end{split}
\end{equation}
a linear combination of the integrals of three complex spherical harmonics
\begin{equation}
\begin{split}
\langle &lm|l_1 m_1|l_2 m_2\rangle = \int d\Omega \left[Y_l^m\right]^* Y_{l_1}^{m_1} Y_{l_2}^{m_2}\\
&= \sqrt{(2l_1+1)(2l_2+1)}(-1)^{m}\begin{pmatrix}
   l & l_1 & l_2 \\
   0 & 0 & 0 
  \end{pmatrix}
  \begin{pmatrix}
   l & l_1 & l_2 \\
   -m & m_1 & m_2 
  \end{pmatrix},\\
\end{split}
\end{equation}
which vanishes unless $l+l_1+l_2=2g$, $g\in \mathbb{Z}$, and $m_1+m_2=m$, thus
truncating the infinite sum over $l$. The integral of three real spherical
harmonics is invariant under all permutations of the order of the functions and
can be simplified into a single complex integral times the appropriate
factors\cite{Homeier1996}. The radial integral of the normalized spherical
Bessel functions $R_{nl}$ is
\begin{equation}
\begin{split}
\label{eq:eris}
&R^l(n_1 l_1, n_2 l_2; n_3 l_3, n_4 l_4) = \\
& \int_0^R dr_1 \int_0^R dr_2 r_1^2 r_2^2 \frac{r_<^l}{r_>^{l+1}}
    R_{n_1l_1}^*(r_1) R_{n_2l_2}^*(r_2) R_{n_3l_3}(r_1) R_{n_4l_4}(r_2) \\
        &= \frac{1}{R} \Bigg[\int_0^1 dx_1 \int_0^{x_1} dx_2 x_1^2 x_2^2 \frac{x_2^l}{x_1^{l+1}}
        u_{n_1l_1}^*(x_1) u_{n_2l_2}^*(x_2) u_{n_3l_3}(x_1) u_{n_4l_4}(x_2) \\
    &\hspace{1em} + \int_0^1 dx_1 \int_{x_1}^{1} dx_2 x_1^2 x_2^2 \frac{x_1^l}{x_2^{l+1}} 
        u_{n_1l_1}^*(x_1) u_{n_2l_2}^*(x_2) u_{n_3l_3}(x_1) u_{n_4l_4}(x_2) \Bigg],
\end{split}
\end{equation}
and can be computed numerically.

\end{document}